\DocumentMetadata{
	lang		= eng-US, 	
	pdfstandard	= A-3u,		
	pdfversion	= 1.7, 		
}

\documentclass[colorlinks,upint,subscriptcorrection,varvw,hyphenate,balance,greek,russian,vietnamese,german]{asmeconf} 

\usepackage{svg}
\usepackage[normalem]{ulem}
\usepackage{natbib}

\hypersetup{%
	pdfauthor={John H. Lienhard},									  
	pdftitle={ASME Conference Paper LaTeX Template},                  
	pdfkeywords={ASME conference paper, LaTeX template, BibTeX style},
	pdfsubject = {Describes the asmeconf LaTeX template},			  
}

\allowdisplaybreaks 

\begin{document}



\ConfName{Proceedings of the ASME 2026\linebreak Fluids Engineering Division Summer Meeting }
\ConfAcronym{FEDSM2026}
\ConfDate{July 26--29, 2026} 
\ConfCity{Bellevue, WA}
\PaperNo{FEDSM2026-184555}

%

\title{Mass-Transfer Control With Microbubbles in Highly Turbulent Decaying Flows} 
 
%
%
%

\SetAuthors{%
    Vivek Kumar\affil{1}, 
	Prasoon Suchandra\affil{1}\CorrespondingAuthor{psuchandra3@gatech.edu},
        Jason Rom\affil{1}, 
	Shivam Prajapati\affil{1}, 
	Suhas Jain\affil{1},  
    Cyrus~Aidun\affil{1}
	}

\SetAffiliation{1}{George W. Woodruff School of Mechanical Engineering, Georgia Institute of Technology, Atlanta, GA 30332, USA\\
Renewable Bioproducts Institute, Georgia Institute of Technology, Atlanta, GA 30332, USA}


\maketitle

\versionfootnote{Documentation for \texttt{asmeconf.cls}: Version~\versionno, \today.}


\keywords{High-Re turbulent bubbly flows, shadowgraph imaging, surfactant-mediated flows}


\begin{abstract}We hypothesize that combining extreme turbulence with a minute reduction in surface tension $\sigma$ (surface tension of the liquid) using surfactant provides a simple and scalable route for controlling micron scale bubble size in gas--liquid systems. To test this, we generate high-intensity turbulence using a multiphase pump [turbulent intensity $\ge 40\%$; Taylor Reynolds number $Re_\lambda=\mathcal{O}(10^3)$; bulk Reynolds number $Re=\mathcal{O}(10^5)$] feeding a straight duct, which produces a decaying turbulent flow where, without additives, bubble coalescence dominates and causes monotonic downstream growth in the mean diameter $d_\mathrm{avg}$ of the bubbles. This growth is governed by the turbulent dissipation rate $\varepsilon$. High-speed imaging, back-lit shadowgraph and particle shadow velocimetry (PSV) quantify bubble statistics ($d_\mathrm{avg}$, and the bubble-size distribution) and turbulence metrics (turbulent kinetic energy $k$, turbulence intensity $\mathcal{I}$, and dissipation rate $\varepsilon$). We then introduce a minute amount ($\sim 0.01\%$ critical micelle concentration) of additive that produces a slight reduction in $\sigma$, used here only as an interfacial tuning knob because the same change in surface tension can be achieved with non surface active agents. This small decrease in $\sigma$ enhances breakup, slightly suppresses coalescence, and makes smaller bubbles more breakup prone, resulting in reduced $d_\mathrm{avg}$ and a narrower bubble-size distribution. Turbulence statistics remain unchanged within experimental uncertainty, indicating that the effect arises entirely from interface rather than hydrodynamic changes. Overall, combining extreme turbulence with a minute reduction in surface tension offers a low complexity and tunable lever for setting bubble-size distributions and intensifying mass transfer in industrial multiphase flows.
\end{abstract}


\begin{nomenclature}
\entry{$D$}{Duct hydraulic diameter}
\entry{$d$}{Bubble diameter}
\entry{$d_g$}{Geometric mean diameter}
\entry{$d_{\rm{H}}$}{Hinze diameter}
\entry{$d_\mathrm{avg}$}{Mean diameter}
\entry{$f(d)$}{Bubble-size distribution}
\entry{$\mathcal{I}$}{Turbulence intensity}
\entry{$K$}{Total number of bins}
\entry{$k$}{Turbulent kinetic energy}
\entry{$\mathcal{L} = x/D$}{Normalized axial position}
\entry{$N$}{Total number of bubbles detected}
\entry{$\rm{Re}$}{Bulk Reynolds number, $VD/\nu$}
\entry{$\rm{Re}_\lambda$}{Taylor-scale Reynolds number, $\mathcal{U}\lambda_T/\nu$}
\entry{$u$}{Instantaneous velocity}
\entry{$\overline{u}$}{Ensemble-averaged velocity}
\entry{$u'$}{Velocity fluctuation}
\entry{$\mathcal{U}$}{Root-mean-square (rms) of velocity fluctuations}
\entry{$V$}{Bulk velocity}
\entry{$\rm{We}$}{Weber number}
\entry{$\Delta$}{PSV vector spacing}
\entry{$\varepsilon$}{Turbulent dissipation rate}
\entry{$\eta$}{Kolmogorov scale}
\entry{$\nu$}{Kinematic viscosity}
\entry{$\phi$}{Void fraction}
\entry{$\Phi$}{Cumulative fraction of bubbles with diameters smaller than $d$}
\entry{$\rho_\ell$}{Liquid density}
\entry{$\sigma$}{Surface tension}
\entry{$\sigma_g$}{Geometric standard deviation (for bubble size)}
\end{nomenclature}


\section{Introduction}

Gas–liquid or multiphase flows occur widely across natural settings and engineered systems, encompassing chemical and biochemical reactors, oil–gas transport pipelines, aeration and wastewater treatment facilities, flotation units, heat exchangers, and heat-pipe pool boiling configurations \citep{delnoij1997computational,kumar2024hydrostatic,kumar2025experimental}. In these flows, dispersed bubbles exert a strong influence on transport phenomena, including mass transfer, mixing efficiency, particle seperation, drag modification, and chemical reaction rates \citep{clift2005bubbles,kumar2023particle}. Bubble behavior in turbulent environments is governed by inherently stochastic processes such as interfacial deformation, liquid-film drainage, and topological transformations driven by coalescence and breakup \citep{MartinezBazan1999,Coulaloglou1977, madanan2021computational}. Under such conditions in high-Reynolds-number flows, turbulent stresses break bubbles into smaller fragments and strongly dictate the subsequent downstream evolution of the two-phase flow, with direct implications for system performance and efficiency~\citep{Kolmogorov1949,Hinze1955,zhang2018investigation}.  Characterizing and controlling the bubble-size distribution (commonly represented using modified Gaussian or log-normal forms) is therefore a central requirement, as it determines the relative population of small and large bubbles and governs the available interfacial area.  Even in the measurement of dynamic surface tension using the bubble pressure tensiometer method~\cite{KUMAR2026139929}.

The statistical distribution of bubble sizes in turbulent flows emerges from a balance between breakup and coalescence events. In theoretical and numerical frameworks, this balance is commonly captured through population-balance formulations coupled to the resolved or modeled turbulence, where size redistribution is governed by breakup and coalescence kernels derived from physical arguments and calibrated empirically \citep{javadi2024bubble,lehr2002bubble,Li2024,Yu2024}. Breakup is initiated when turbulent inertial forces exceed stabilizing surface tension, resulting in fragmentation above a threshold Weber number \citep{Kolmogorov1949,Hinze1955,ni2024deformation}. In turbulence, the Weber number can be expressed as
\begin{equation} \label{eq:Webernumber}
\mathrm{We} \;=\; 2.13\,\frac{\rho_\ell}{\sigma}\,(\varepsilon d)^{2/3}
\end{equation}
where $d$ is the bubble diameter, $\varepsilon$ represents the local turbulent dissipation rate, $\sigma$ is the interfacial tension, and $\rho_\ell$ is the liquid density~\citep{ni2024deformation}. The largest bubble that can persist without breaking corresponds to the critical Weber number, $\mathrm{We}_{\mathrm{crit}}$, and defines the Hinze diameter, $d_{\mathrm{H}}$ \citep{Hinze1955,hesketh1987bubble}.  The surface tension of the working liquid is $\sigma = 72\,\mathrm{mN/m}$, and the critical Weber number is $\mathrm{We}_{\mathrm{crit}} \approx 1.1$. The presence of surfactants lowers the effective interfacial tension, thereby reducing the Hinze scale and rendering bubbles more susceptible to turbulent fragmentation. As a result, breakup can occur at lower turbulent dissipation levels, shifting the bubble-size distribution toward smaller diameters. The presence of surfactants modifies bubble-bubble interactions by altering interfacial mobility and film stability. Although elevated turbulence promotes frequent collisions, the associated strong shear and unsteady velocity gradients significantly reduce the effective contact time during bubble encounters. When surfactants are present, liquid-film drainage is further slowed by reduced surface tension and surfactant-induced Marangoni stresses, which stabilize the interfacial film and suppress coalescence even under frequent collisions \citep{Liao2010,hagesaether2000coalescence}. In addition, wake-induced entrainment by larger bubbles generates localized flow structures that enhance spatial heterogeneity in bubble concentration and collision dynamics. As a consequence, highly turbulent, surfactant-laden flows often exhibit a regime in which collision frequency is high but net coalescence remains limited, emphasizing the complex and competing roles of turbulence and interfacial physics in determining bubble population evolution \citep{lance1991turbulence,Liao2010,hagesaether2000coalescence,Laakkonen2007}. These observations underscore the need to extend traditional coalescence models to explicitly account for the coupled effects of turbulence intensity, interfacial modification by surfactants, and wake-driven flow structures in high-Reynolds-number bubbly flows \citep{Laakkonen2007}.

Despite their widespread use, existing coalescence models face fundamental challenges when applied to the extreme and rapidly evolving turbulence encountered downstream of pumps and injectors, where bulk Reynolds numbers reach $\mathrm{Re}=V D/\nu \sim 10^5$ and Taylor-scale Reynolds numbers approach $\mathrm{Re}_\lambda=\mathcal{U}\lambda_T/\nu \sim 10^3$. Here, $V$ denotes the bulk mean velocity, $D$ the hydraulic diameter of the duct, $\nu$ the kinematic viscosity, $\mathcal{U}$ the root-mean-square of velocity fluctuations, and $\lambda_T$ the Taylor microscale. Such flows are marked by large inlet turbulence intensities (approaching 50\%), followed by strong axial decay \citep{javadi2026large}. Classical population-balance closures, including those of Coulaloglou and Tavlarides and Prince and Blanch, assume locally homogeneous and quasi-steady turbulence and employ drainage-based efficiencies calibrated primarily under moderate conditions. Consequently, these models are unable to represent the sharp spatial gradients in turbulence intensity and the rapid downstream evolution of bubble-size distributions in developing bubbly flows.  Experimental data under such extreme, decaying, and surfactant-influenced conditions are scarce, with most prior studies limited to weak or moderate turbulence, homogeneous grid turbulence, or pseudo-turbulent bubble swarms \citep{bouche2012homogeneous,bouche2014homogeneous}.

The present study provides novel experimental insight into bubble dynamics in a square duct subjected to extreme pump-driven turbulence in the presence of surfactants, with bulk Reynolds numbers exceeding $10^5$ and $\mathrm{Re}_\lambda$ approaching 900. In this configuration, very high inlet turbulence intensity, with $\mathcal{I}\gtrsim 30\%$, fragments incoming bubbles to sizes near the classical Hinze scale, reflecting a balance between turbulent breakup and suppressed coalescence~\citep{Hinze1955,MartinezBazan1999}. As the flow convects downstream, the turbulent dissipation rate decays rapidly along the axial direction. Since the Hinze scale follows $d_{\mathrm{H}}\propto\varepsilon^{-2/5}$, it increases as turbulence weakens, progressively relaxing the breakup constraint~\citep{kumar2025bubblejfm}. However, the evolution of bubble size is not dictated solely by turbulence decay but also depends on interfacial properties, void fraction, and the initial bubble-size distribution~\citep{Prince1990}. As $\varepsilon(t)$ decreases, collision frequencies remain elevated due to residual turbulence, while surfactant-stabilized interfaces favor coalescence, driving a gradual transition toward coalescence-dominated growth. Nevertheless, the Hinze scale continues to impose an upper bound on bubble size, as bubbles exceeding $d_{\mathrm{H}}$ remain susceptible to breakup. This interplay produces a sustained regime in which pump-induced fragmentation near the inlet gives way to coalescence-driven bubble growth downstream. Such a strongly coalescent regime under decaying, high-Reynolds-number, surfactant-modified turbulence has not been previously quantified~\citep{ruth2022experimental}.

\section{Methodology} \label{section:measuring technique}

\subsection{Experimental setup}

Our experiments are conducted in a multiphase flow loop with a vertically oriented square duct test-section, as shown in figure \ref{fig:exp_setup}. Bubble generation is achieved using an inline stainless-steel sparger, which injects compressed air into the liquid stream to produce a fine initial dispersion. The resulting two-phase mixture then passes through a regenerative turbine pump, which is the primary source of turbulence generation. Once the air is injected in the liquid line, the air-water mixture goes through a regenerative pump which produces a highly turbulent flow with very fine bubbles\cite{javadi2026large}. This configuration produces a highly turbulent flow that subsequently decays along the downstream test section. The combined design ensures controlled bubble injection and well-defined turbulent conditions for systematic measurements~\cite{kumar2025bubblejfm}. More details of this experimental facility can be found in Kumar et al. \cite{kumar2025bubblejfm,suchandra2025surfactant}.

\begin{figure*}[htbp]
    \centering
    \includegraphics[width=0.96\textwidth]{Images_exp/exp_setup_v1.png}
    \caption{Schematic of the experimental setup; also showing the components of the imaging system, raw and processed back-lit shadowgraph bubble images, and the steps of particle shadow velocimetry.}
    \label{fig:exp_setup}
\end{figure*}

\subsection{Back-lit shadowgraph for bubble imaging}

A high-speed imaging system is utilized to capture bubble dynamics within the designated observation sections (see figure \ref{fig:exp_setup}). The setup features a high-speed camera and a halogen backlight, ensuring uniform and high-contrast visualization of the bubbles. Imaging is performed at a shutter speed of 3.33\,$\mu$s, providing a spatial resolution of 5$\pm$0.2\,$\mu$m. Image sequences are recorded in lossless \textit{.tif} format at 1\,frame per second\,(fps) intervals, which both maximized image fidelity and minimized the likelihood of repeated bubble detection during subsequent post-processing.  Further details of the experimental facility, including the bubble generation method and the turbulence generation mechanism, can be found in our previous study~\cite{kumar2025bubblejfm}.

The imaging system is calibrated using a reference wire strand of known diameter, and the resulting spatial resolution is 5\,$\pm 0.2\, \mu$m/pixel. Raw bubble images are first preprocessed through background subtraction to suppress noise and enhance bubble edges (see figure \ref{fig:exp_setup}). Bubble contours  are detected using the Canny edge detection method \citep{rong2014improved}, and watershed segmentation \citep{seal2015watershed} is applied to resolve overlapping bubbles. Furthermore, the bubble-size distribution for select images is cross-validated using the open-source, deep learning-based automated bubble detection algorithm developed by Kim \& Park \citep{kim2021deep} an in house code~\citep{nayak2026depth}. The projected two-dimensional area $A$ of each bubble is obtained from the shadowgraph images, using the area-equivalent diameter, i.e., $d = d_{\text{equiv}} = \left( \frac{4A}{\pi} \right)^{1/2}$.

In these cases 20 images were analysed. In future, we plan to expand to the data set for stastical converagnce. The Sauter mean diameter ( $d_\mathrm{avg}$ ) is computed as:
\[
d_\mathrm{avg} = \frac{\sum_{i=1}^{N} d_i}{N}.
\]
\noindent where $d_i$ is bubble diameter, and \( N \) is the total number of bubbles detected. To assess uncertainty in \( d_\mathrm{avg} \), a sensitivity analysis is performed by varying segmentation thresholds and repeating the analysis on selected datasets.  Using an area-equivalent diameter implicitly assumes approximate axisymmetry of the projected bubble shape, which is justified here because the bubbles satisfy $\mathrm{We} \lesssim 1$ and they remain nearly spherical.

\subsection{Particle shadow velocimetry for turbulent field}

Particle shadow velocimetry (PSV) \citep{estevadeordal2005,khodaparast2013,hessenkemper2018,jassal2025particle}, a cost-effective alternative to particle image velocimetry (PIV), is implemented to measure primarily the liquid-phase velocity in the vertical duct. As with bubble imaging, a high-power light source illuminates the flow from behind (see figure \ref{fig:exp_setup}), casting sharp shadows of flow tracers onto the sensor of a high-speed camera operated at 20000 fps. The depth of field of the lens used is $\approx0.5$\,mm, which is comparable to the typical laser-sheet thickness in planar PIV applications. The flow was seeded with nearly neutrally buoyant hollow glass spheres (mean diameter $\sim$ 10\,$\mu$m) which serve as PSV flow tracers. The response time of these tracers/particles is much smaller than the characteristic flow time scale, i.e. Stokes number\,$\ll$\,1, ensuring faithful tracking. The calibrated field of view covered roughly 3\,mm in the vertical direction at approximately 10\,$\mu$m/pixel resolution. Note that the image resolution for PSV is lower than that for bubble imaging discussed previously. This is to capture sufficient particle and small bubble motions within the field of view. The PSV images (with both particle and bubble shadows) are processed, subtracting background and removing large bubble shadows, to convert them into formats suitable for cross-correlation based PIV processing. Image pairs are then processed with the open-source software OpenPIV~\citep{OpenPIV}. We used 128 px $\times$ 128 px interrogation windows with 75\% overlap. The resulting instantaneous velocity fields are further corrected using the universal outlier detection method of Westerweel \& Scarano \cite{Westerweel_2005}. We conservatively used 1000 image-pairs to get converged velocity statistics. The controlled nature and repeatability of our experiments and analyses are checked via obtaining converged turbulence statistics from multiple independent runs under identical operating conditions. The uncertainties in velocity and turbulence statistics are estimated using the methods of Wieneke \cite{Wieneke_2015} and Sciacchitano \& Wieneke \cite{Sciacchitano_2016}.

From the validated vector fields, we compute ensemble-averaged and fluctuation quantities, following the Reynolds decomposition of the instantaneous velocity $u = \overline{u} + u'$. The two-component turbulent kinetic energy ($k$) and turbulence intensity ($\mathcal{I}$) are estimated from the instantaneous fluctuations ($u'$), as shown below in equations \ref{eq:tke_full} and \ref{eq:TI}: \\

\begin{equation}\label{eq:tke_full}
k = \frac{1}{2}\left(\overline{u'^2} + \overline{v'^2} + \overline{w'^2}\right) \approx \frac{1}{2}\left(\overline{u'^2} + 2 \ \overline{v'^2}\right) 
\end{equation}

\noindent where $\overline{w'^2} \approx \overline{v'^2}$ is a reasonable assumption given the axisymmetry of the flow geometry \citep{gavrilakis1992squareduct,owolabi2016squareduct}.  The turbulence intensity $\mathcal{I}$ can be evaluated as,


\begin{equation}\label{eq:TI}
\mathcal{I} = \frac{\mathcal{U}}{V} = \frac{\sqrt{\frac{2\,k}{3}}}{V}
\end{equation}

\noindent where $V$ is the bulk velocity and $\mathcal{U}$ denotes the root-mean-square (rms) value of velocity fluctuations.

\begin{figure}[htbp]
    \centering
    \includegraphics[width=0.4\textwidth]{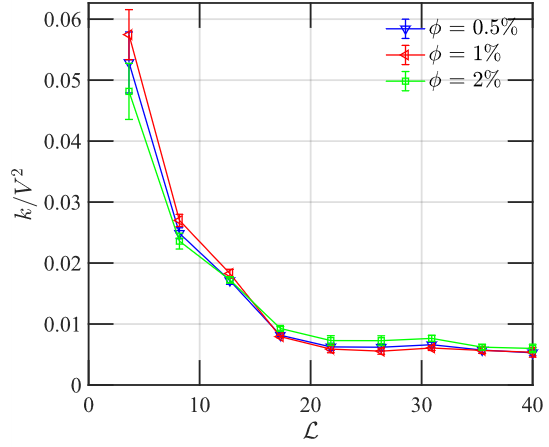}
    \caption{Streamwise evolution of nomralised turbulent kinetic energy ($k/V^2$) at different $\phi$.}
    \label{fig:tke_decay}
\end{figure}

The  measured dissipation rate ($\varepsilon_m$) is first calculated using the 2D velocity gradients, via 
\begin{equation} \label{eq:dissipation}
\varepsilon_{\mathrm{m,axi}} = \nu \overline{\left( 
- \left( \frac{\partial u'}{\partial x} \right)^2 
+ 2 \left( \frac{\partial v'}{\partial x} \right)^2 
+ 2 \left( \frac{\partial u'}{\partial y} \right)^2 
+ 8 \left( \frac{\partial v'}{\partial y} \right)^2  
\right)}
\end{equation}

\noindent under the local axisymmetry assumption \citep{george1991locally,xu2013accurate} which is appropriate for our flow geometry. Due to limited resolution of PSV used here, the dissipation rates obtained from measured velocity fields and velocity gradients are considerably underestimated \citep{Lavoie2007,xu2013accurate,Verwey2022}. For the current work, we use the direct numerical simulations (DNS) calibrated, structure function and spectra based empirical method of Xu \& Chen \cite{xu2013accurate} which corrects the measured dissipation rate $\varepsilon_{\mathrm{m}}$ as a function of $\Delta/\eta$ and estimates the corrected dissipation rate $\varepsilon_{\mathrm{cor}}$, as
\[
\frac{\varepsilon_{\mathrm{m}}}{\varepsilon_{\mathrm{cor}}}
=
a_1 e^{a_2(\Delta/\eta)} + b_1 e^{b_2(\Delta/\eta)}
\]
\noindent where $\Delta$ is the PSV vector spacing, $\eta$ is the Kolmogorov scale, and  $a_1=1.1910$, $a_2=-0.03081$, $b_1=-0.1835$, $b_2=-0.2062$ for a top-hat interrogation window. Independently, the corrected dissipation is verified against the classical homogeneous isotropic turbulence (HIT) scaling ($\varepsilon \sim C_\varepsilon u'^3/\ell$, with relative ratio of $C_\varepsilon \approx \mathcal{O}(1)$ and $\ell \sim$ hydraulic diameter $D$), and the resulting values (order of magnitude) fairly agree, as expected at high Reynolds numbers.

\begin{figure*}[htbp]
    \centering
    \includegraphics[width=0.65\textwidth]{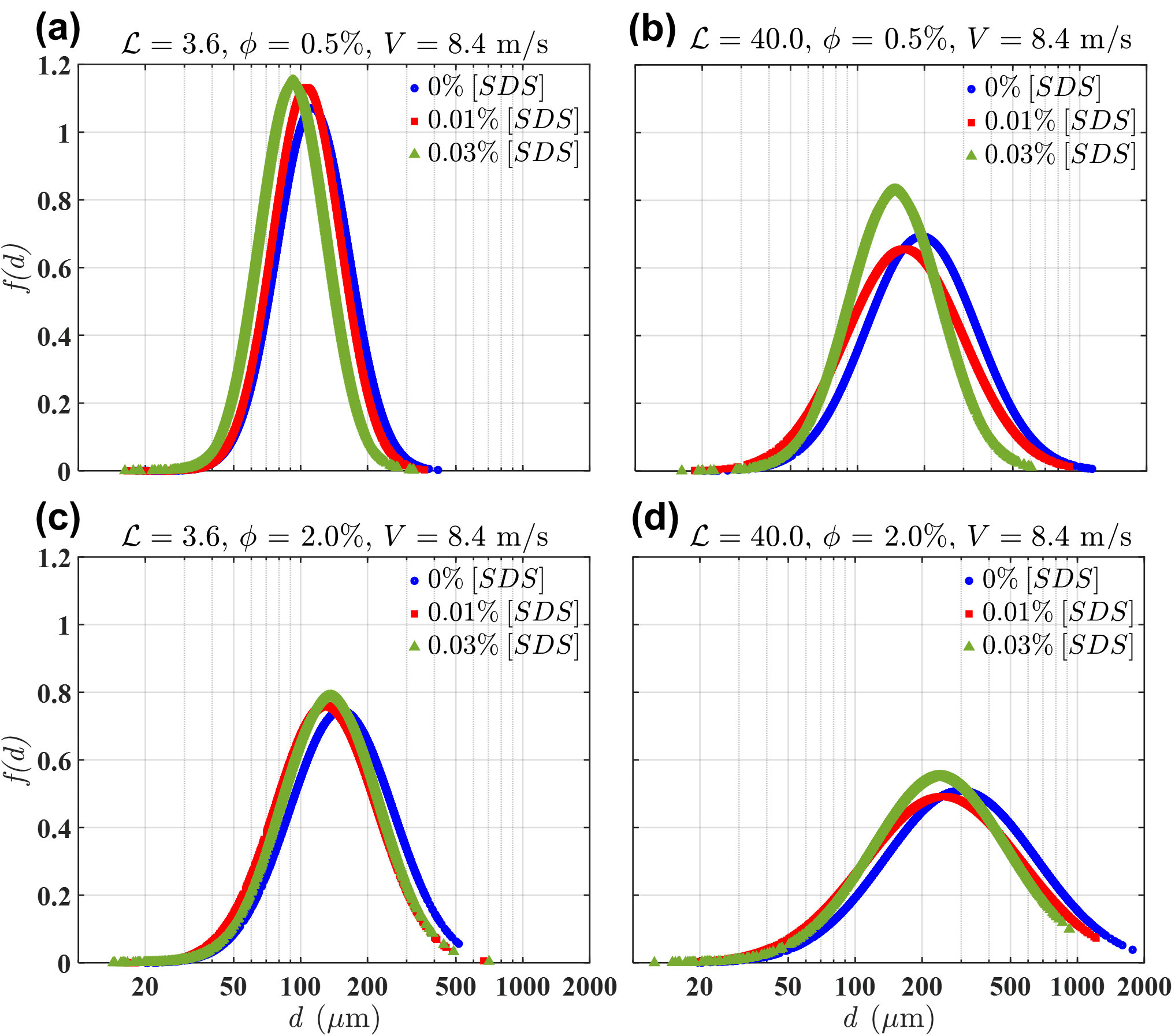}
    \caption{The bubble-size distributions ($f(d)$) at two streamwise locations, $\mathcal{L} = x/D = 3.6$ and $\mathcal{L} = 40.0$, for two void fractions $\phi =$  0.68\% $\approx$  0.5\% and 2.5\% $\approx$  2.0\%. Each sub-figure has plots for the three SDS concentrations, 0\%, 0.01\% and 0.03\% of CMC.}
    \label{fig:BSDs}
\end{figure*}

\begin{figure*}[htbp]
    \centering
    \includegraphics[width=0.67\textwidth]{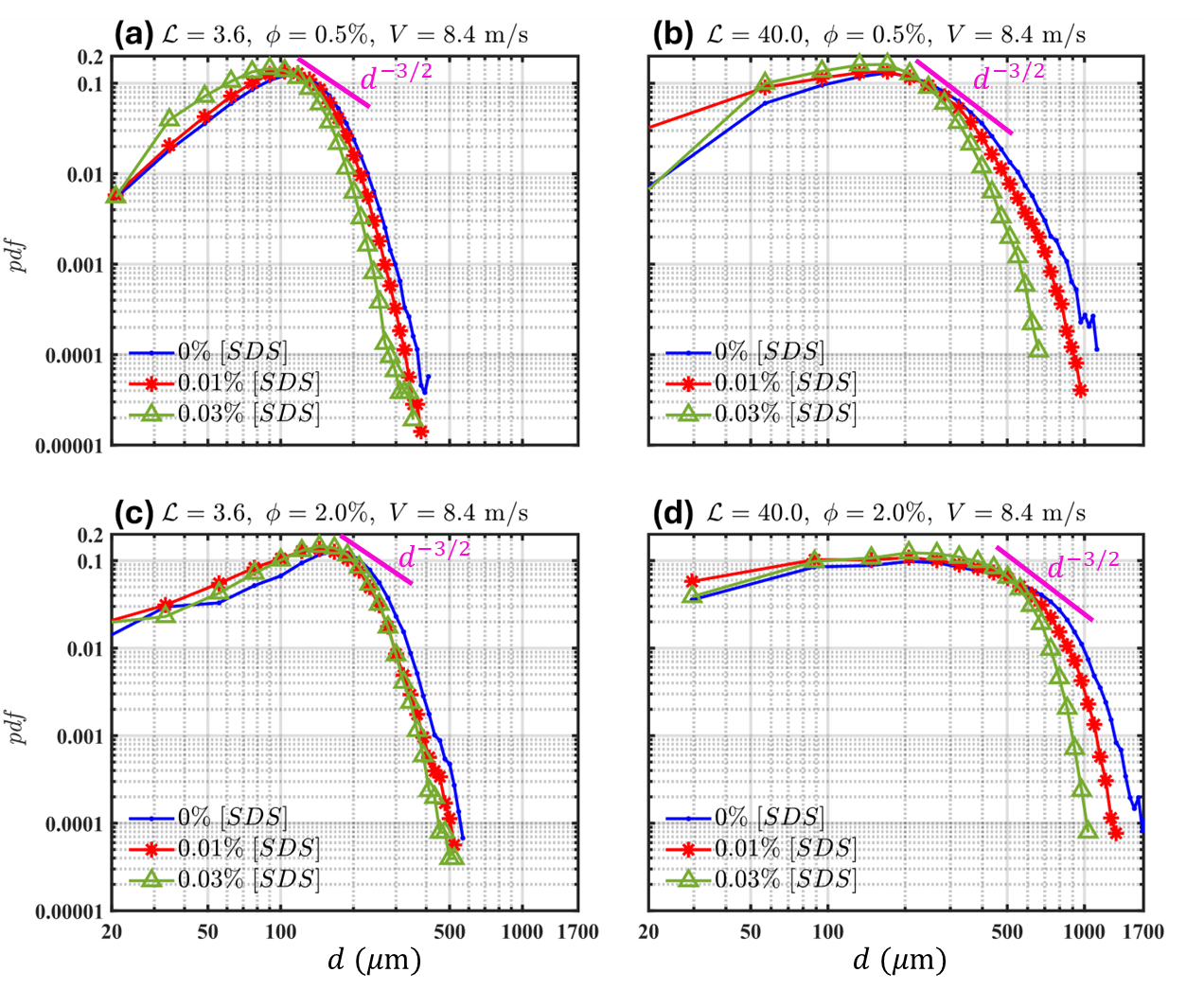}
    \caption{$pdf$ of bubble sizes at two streamwise locations, $\mathcal{L} = x/D = 3.6$ and $\mathcal{L} = 40.0$, for two void fractions $\phi =$  0.68\% $\approx$  0.5\% and 2.5\% $\approx$  2.0\%. Each sub-figure has plots for the three SDS concentrations, 0\%, 0.01\% and 0.03\% of CMC.}
    \label{fig:PDFs}
\end{figure*}

\section{Results and discussion}

In this paper, we present results from experiments conducted at a single bulk velocity of $V = 8.4$ m/s ($\mathrm{Re} = 1.3 \times 10^5$), at three void fractions $\phi =$  0.68\% $\approx$  0.5\%,  1.18\% $\approx$  1\% and 2.5\% $\approx$  2.0\%.  The axial decay of turbulent kinetic energy (or turbulence intensity), reaching up to approximately $90\%$, as shown in figure~\ref{fig:tke_decay}), which characterizes the strongly decaying turbulence downstream of the pump. The measurements are made near the center plane of the duct. We look at the effect of addition of surfactant sodium dodecyl sulphate (SDS) at 0\%, 0.01\% and 0.03\% of critical micelle concentration (CMC). The bubble-size distributions ($f(d)$) and the probability density functions ($pdf$) of bubble sizes are presented at two streamwise locations, $\mathcal{L} = x/D = 3.6$ and $\mathcal{L} = 40.0$ (far upstream and far downstream locations), for two void fractions $\phi =$  0.68\% $\approx$  0.5\% and 2.5\% $\approx$  2.0\%. The streamwise evolutions of the bubble size (represented using mean diameter $d_\mathrm{avg}$) are also presented, for all three void fractions $\phi =$  0.68\% $\approx$  0.5\%,  1.18\% $\approx$  1\% and 2.5\% $\approx$  2.0\%.

\subsection{Bubble-size distribution and probability distribution function}

The modified Gaussian function [$f(d)$] \citep{friedlander2000smoke} is used here to characterize bubble-size distribution, given by

\begin{equation*} \label{eq:f(d)}
    f(d) = \frac{\partial \Phi}{\partial \ln d} = \frac{1}{\sqrt{2\pi} \ln \sigma_g} 
\exp \left[ -\frac{1}{2} \left( \frac{\ln(d/d_g)}{\ln \sigma_g} \right)^2 \right]
\end{equation*}

\begin{equation*} \label{eq:d_g and sigma_g}
\text{with} \, d_g = \left( \prod_{i=1}^{K} (d_i)^{n_i} \right)^{\frac{1}{K}} \, , \, \ln \sigma_g = \sqrt{ \frac{ \sum_{i=1}^{K} n_i (\ln d_i - \ln d_g)^2 }{N} } 
\end{equation*}

\noindent where $d_i$ is the representative diameter for bin $i$, $n_i$ is the number of bubbles in that bin, $K$ is the total number of bins, $d_g$ and $\sigma_g$ are the geometric mean diameter and geometric standard deviation \citep{razzaque2003bubble}, and $\Phi$ is the cumulative fraction of bubbles with diameters smaller than $d$. These bubble-size distributions are shown in figure \ref{fig:BSDs}. We see sharp-peaked bubble distributions, with small mean diameters, near the inlet because of strong turbulence resulting in fine bubbles. 
As we move downstream, comparing the sub-figures on left vs right, fewer new micro-bubbles are formed due to decaying turbulence, and the bubbles start to coalesce or merge, leading to the distribution becoming broad and moving toward larger diameters. The axial decay of turbulent kinetic energy (or turbulence intensity), reaching up to approximately $90\%$, has been reported in our previous work~\cite{kumar2025bubblejfm}, which characterizes the strongly decaying turbulence downstream of the pump.  Near the inlet, the Weber number estimated using equation (1) ranges from approximately $0.1$ to $1.5$. As the turbulence decays downstream, the corresponding Weber number changes to approximately $0.05$--$0.9$ near the outlet. With increasing void fractions, comparing the sub-figures on top vs bottom, we see the distribution to broaden and shift toward larger diameters. This is due to increased bubble-bubble collisions at higher gas concentrations, leading to bubble coalescence. 

From figure \ref{fig:BSDs}, we also see that the addition of surfactant shifts the overall bubb-size distribution and its peak to smaller diameters, and this expected observation is more apparent for the lower void fraction case ($\phi =$  0.68\% $\approx$  0.5\%). The addition of surfactant also generally causes the distributions to become narrower. For the higher void fraction case ($\phi =$ 2.5\% $\approx$  2.0\%), the addition of surfactant does shift the distribution to lower diameters but the difference between the two non-zero surfactant concentration cases is not that clearly demarcated. Even though the flow remains coalescence dominated, there is slight suppression of coalescence with addition of surfactant.


As with $f(d)$, the $pdf$'s are presented in figure \ref{fig:PDFs}, with bubble diameters $d$  (dissipation rate $\varepsilon$ contributes to the estimation of Hinze scale $d_{\mathrm{H}}$; see equation \ref{eq:Webernumber}). The plot shows $d^{-3/2}$ scaling (for sub Hinze scale bubbles, not shown) which indicates no active breakup implying coalescence dynamics. The effect of addition of surfactant in shifting the bubble distribution to smaller diameters is clearly seen here. The difference between all the surfactant cases is clearly demarcated.

\subsection{Streamwise evolution of bubble size}

\begin{figure}[htbp]
    \centering
    \includegraphics[width=0.4\textwidth]{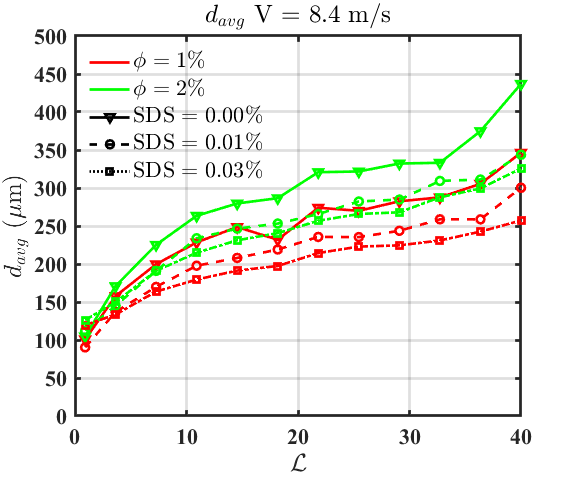}
    \caption{Streamwise evolution of  mean diameter $d_\mathrm{avg}$, for three void fractions $\phi =$  0.68\% $\approx$  0.5\%,  1.18\% $\approx$  1\% and 2.5\% $\approx$  2.0\%, and surfactant SDS concentrations of 0\%, 0.01\% and 0.03\% of CMC.}
    \label{fig:bub_evol}
\end{figure}
The effect of surfactants in suppressing bubble coalescence is clearly evident in Fig.~\ref{fig:bub_evol} through the streamwise evolution of the  mean diameter ($d_\mathrm{avg}$). As new interfacial area is created, surfactant molecules redistribute between the interface and the bulk phase, thereby modifying the local interfacial surface tension. This redistribution stabilizes the thin liquid film formed between approaching bubbles and inhibits coalescence. Consequently, a clear reduction in bubble growth is observed when comparing the surfactant-free cases (solid curves) with the surfactant-laden cases (dashed and dotted curves). The systematic decrease in bubble size with increasing surfactant concentration is apparent, with higher concentrations leading to a monotonic reduction in the Sauter mean diameter along the test section.


\section{Conclusions}

 To study highly turbulent surfactant-laden bubbly flow, we generate high-intensity turbulence using a multiphase pump [Taylor Reynolds number $\rm{Re}_\lambda=\mathcal{O}(10^3)$; bulk Reynolds number $\rm{Re}=\mathcal{O}(10^5)$] feeding a straight duct, which produces a decaying turbulent flow where, without additives, bubble coalescence dominates and causes monotonic downstream growth of bubbles. Three void fractions,  0.68\% $\approx$  0.5\%,  1.18\% $\approx$  1\% and 2.5\% $\approx$  2.0\%, are studied. High-speed imaging, back-lit shadowgraph and particle shadow velocimetry are used to quantify bubble size statistics and turbulence metrics. We then introduce a small amount of surfactant (as sodium dodecyl sulphate at 0\%, 0.01\% and 0.03\% of critical micelle concentration) that produces a slight reduction in surface tension, suppressing coalescence, and making smaller bubbles more breakup prone, resulting in reduced bubble diameters and a narrower bubble-size distribution. Turbulence statistics remain unchanged within experimental uncertainty. We saw that combining extreme turbulence with a minute reduction in surface tension offers a simple and tunable way to set bubble-size distributions and modulate mass transfer in multiphase flows. In future, its interseting to explore similar bubbly flow in Non Newtonian decaying flows~\cite{prajapati2026laminar,prajapati2025laminar}. Additionally, measurements are being conducted at higher resolutions to resolve smaller bubble diameters. Also, increasing the number of images further improves statistical convergence, as we can see those convergance is not good yet.


\section*{Acknowledgments}

The information, data, or work presented herein was funded in part by the Advanced Research Projects Agency-Energy (ARPA-E), U.S. Department of Energy, under Award Number DE-AR0001587. The views and opinions of authors expressed herein do not necessarily state or reflect those of the United States Government or any agency thereof. In disclosure, the part of this work was presented at APS-DFD. 


\bibliographystyle{asmeconf}  
\bibliography{asmeconf-sample}

\appendix


\end{document}